\documentclass[preprint,aps]{revtex4}

\usepackage{graphicx}
\usepackage{dcolumn}
\usepackage{bm}

\begin{document}


\title{Doppler-free spectroscopy of the $^{1}S_{0}$--$^{3}P_{0}$ optical clock transition in laser-cooled fermionic isotopes of neutral mercury}

\author{M. Petersen}
\author{R. Chicireanu}
\author{S.T. Dawkins}
\author{D.V. Magalh\~{a}es}
\altaffiliation[Also at ]{Instituto de F\'{\i}sica de S\~{a}o Carlos, USP-PO
Box 369, 13560-970, S\~{a}o Carlos, SP, Brazil.}
\author{C. Mandache}
\altaffiliation[Also at ]{National Institute for Laser Physics,
Plasmas and Radiation, Plasmas and Nuclear Fusion Laboratory,
Bucharest, Magurele, PO Box MG 7, Romania.}
\author{Y. Lecoq}
\author{A. Clairon}
\author{S. Bize}

\affiliation{%
LNE-SYRTE, Observatoire de Paris\\
75014 Paris, France. }

\date{\today}

\begin{abstract}
We have performed for the first time direct laser spectroscopy of the
$^{1}S_{0}$--$^{3}P_{0}$ optical clock transition at 265.6~nm in
fermionic isotopes of neutral mercury laser-cooled in a
magneto-optical trap. Spectroscopy is performed by measuring the
depletion of the magneto-optical trap induced by the excitation of
the long-lived $^{3}P_{0}$ state by a probe at 265.6~nm.
Measurements resolve the Doppler-free recoil doublet allowing for a
determination of the transition frequency to an uncertainty well
below the Doppler-broadened linewidth. We have performed absolute
 measurement of the frequency with respect to an
ultra-stable reference monitored by LNE-SYRTE fountain primary frequency
standards using a femtosecond laser frequency comb. The measured
frequency is $1128575290808\pm 5.6$~kHz in $^{199}$Hg and
$1128569561140\pm 5.3$~kHz in $^{201}$Hg, more than 4 orders of magnitude better
than previous indirect determinations. Owing to a low sensitivity to
blackbody radiation, mercury is a promising candidate for
reaching the ultimate performance of optical lattice clocks.

\end{abstract}

\pacs{32.30.Jc, 06.30.Ft, 42.62.Fi, 37.10.-x}
\maketitle

The performance of optical atomic clocks is improving at a high pace.
Optical clocks are now surpassing atomic fountain clocks based on
microwave transitions
\cite{Rosenband2008,Boyd2007,Ludlow2008,Schneider2005}. Some of
these optical transitions are now recognized as secondary
representations of the unit of time of the international system of
units (SI) opening the way to a new definition of the SI second
based on an optical transition in the coming years. Atomic clocks
also represents a powerful tool for testing fundamental physical laws.
For instance, stability of natural constants and thereby that of
fundamental interactions (electro-weak, strong interaction) can be
tested to high levels of precision, providing constraints that are
independent of any assumption related to cosmological models
\cite{Marion2003,Fischer2004,Blatt2008,Fortier2007,Peik2004,Rosenband2008}.
Such tests provide precious experimental information to help the
search for unified theories of fundamental interactions.

Optical lattice clocks using strontium atoms have now demonstrated accuracies at
the $10^{-16}$ level \cite{Ludlow2008}, a factor of $\sim 4$ better
than the best atomic fountains but still a factor of $\sim 4$ worse than
the best optical single ion clock based on Al$^{+}$
\cite{Rosenband2008}. At this level of uncertainty, the blackbody
radiation shift is the largest correction and the largest
contribution to the strontium clock uncertainty. In future
development, the blackbody radiation shift will remain a severe
limitation to the accuracy at the $10^{-17}$ level. An optical clock
using ytterbium \cite{Poli2008} will have the same limitation since
the blackbody shift is no more than a factor of 2 smaller in fractional
terms \cite{Porsev2006}. In contrast, neutral mercury has been
recognized as having a low sensitivity to blackbody radiation
\cite{Palchikov2004,Petersen2007,Hachisu2008} whilst retaining all
other desirable features for an optical lattice clock. Mercury has
the potential to achieve uncertainty in the low $10^{-18}$ and therefore to
compete with the best single ion optical clocks
\cite{Rosenband2008}. Mercury is also an interesting candidate in
the search for variations of natural constants owing to its
relatively high sensitivity to variations of the fine structure
constant \cite{Angstmann2004}. However, laser cooling of neutral
mercury has been pursued and achieved only recently
\cite{Petersen2008,Hachisu2008} due to the challenging requirement of
 deep-UV laser sources.

In this paper, we report the first direct laser spectroscopy of the
$^1S_0$--$^3P_0$ clock transition at $265.6$~nm in the two naturally occurring fermionic
 isotopes of mercury. Spectroscopy is performed on a sample
of cold atoms released from a magneto-optical trap (MOT). We this approach, we resolve
 the Doppler-free recoil doublet allowing for a
determination of the transition frequency with an uncertainty well
under the Doppler-broadened linewidth. Absolute measurement of the
frequency is performed using an optical frequency comb.

\begin{figure}[h]
\includegraphics[width= 6cm]{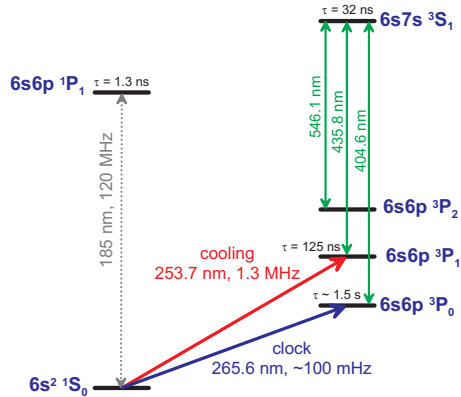}
\caption{\label{fig:Hg_levels} Relevant energy levels of mercury.
The $^1S_0-^3P_1$ transition at 253.7~nm is used for magneto-optical
trapping. The clock transition studied in this paper is the
$^1S_0-^3P_0$ transition at 265.6~nm. The mentioned natural
linewidth of $\sim 100$~mHz is for $^{199}$Hg based on
\cite{Bigeon1967}.}
\end{figure}

Figure \ref{fig:Hg_levels} shows the low-lying energy levels of
mercury. Mercury has an alkaline-earth like electronic structure similar to
those of strontium or ytterbium. In our experiment, laser cooling of
mercury is achieved using the $^1S_0$--$^3P_1$ transition at
253.7~nm with a natural linewidth of $1.3$~MHz. Cooling light is
provided by quadrupling a Yb:YAG thin disk laser delivering up to
7~W of single frequency light at 1014.8~nm. A commercially available
doubling stage using a temperature tuned LBO crystal within a
bow-tie configuration build-up cavity generates up to 3~W of power
at 507.4~nm. A second doubling stage uses a 90$^{\mathrm{o}}$-cut
anti-reflection coated 7~mm long angle tuned BBO, also in a bow-tie
configuration.
Out-coupling of the second harmonic at 253.7~nm is achieved by using
an harmonic separator mirror as one of the build-up cavity mirrors.
Up to 800~mW of CW power have been generated with this system. In
practice, the system is set to generate 100 to 150~mW to avoid the
rapid degradation of the harmonic separator observed at higher
output power. The frequency of this  light is stabilized
to the saturated absorption feature observed in a room temperature
mercury vapor cell with a 1~mm interaction length. The residual
jitter is estimated to be less than $100$~kHz which is suitably
small compared to the $1.3$~MHz natural linewidth of the cooling
transition.

A 2 dimensional magneto-optical trap (2D-MOT) geometry
\cite{Dieckmann1998} is used to generate a slow atom beam which in
turn is used to load a conventional MOT. A vapor pressure of $\sim
2\times 10^{-7}$~mbar of mercury is kept in the 2D-MOT chamber by
cooling a few grams of mercury held in a copper bowl down to
$-55^{\mathrm{o}}$C using two stage Peltier element inside the
vacuum chamber. The 2D-MOT is formed at the intersection of two
orthogonal pairs of $\sigma^{+}$--$\sigma^{-}$ polarized
retro-reflected beams with a total power of $\sim 50$~mW. The
longitudinal and transverse diameters at $1/e^2$ are $\sim 10$~mm
and $\sim 8$~mm. The intersection overlaps with the center of a
2-dimensional quadrupole magnetic field with a gradient of
$0.2$~mT.mm$^{-1}$ generated by 4 rectangularly shaped coils located
outside the vacuum chamber. A $1.5$~mm diameter and $10$~mm long
hole in the 2D-MOT chamber allows the beam of slow atoms confined at
the center of the quadrupole field through, while providing high
differential pumping between the 2D-MOT and the MOT chambers. The
slow atom beam is directed toward the center of the MOT, $70$~mm
away from the output of the 2D-MOT. The MOT is generated at the
intersection of three orthogonal pairs of retro-reflected $\sigma^{+}-\sigma^{-}$
polarized laser beams with a diameter of $\sim 6.6$~mm and a power
of $\sim 15$~mW each. Coils located outside the vacuum chamber
generate the magnetic quadrupole field with a gradient of
$0.15$~mT.mm$^{-1}$ along the strong axis. Our present setup forces
the detuning of the 2D-MOT and the MOT to be the same. We find that
a red detuning of $-5.5$~MHz corresponding to $-4.4~\Gamma$
optimizes the number of atoms in the MOT. Detection of the MOT is
performed by collecting fluorescence light onto a low noise
photodiode. Measurements with the most abundant $^{202}$Hg isotope
indicate that $\sim 5\times 10^{6}$ atoms are captured with a
loading time constant of $2.3$~s. Further measurements of atom
number based on absorption of a weak 253.7~nm probe are in
agreement. $^{199}$Hg and $^{201}$Hg isotopes typically show
slightly reduced atom numbers.

\begin{figure}[ht]
\includegraphics[width= \linewidth]{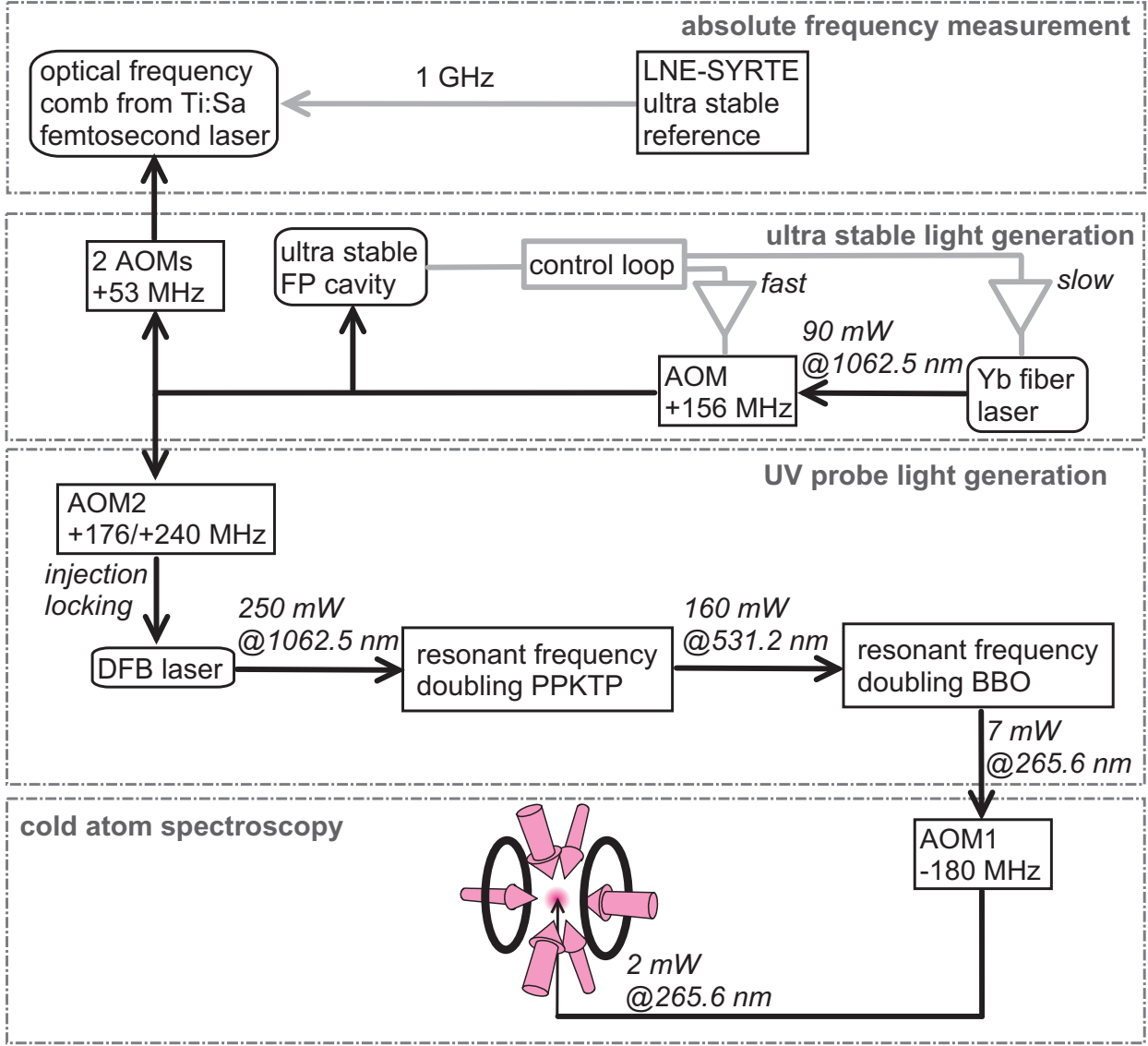}
\caption{\label{fig:probe_light} Generation of the probe light at
265.6~nm and referencing to primary frequency reference. AOM2
frequency depends on the isotope.}
\end{figure}

As shown in fig. \ref{fig:probe_light}, laser light at 265.6~nm for
probing the $^1S_0-^3P_0$ clock transition is generated by frequency
quadrupling the 1062.5~nm output of a distributed feedback (DFB)
laser diode delivering $250$~mW of useful light. The first doubling
is accomplished with $64\%$ efficiency using a periodically-poled
KTP crystal and a bow-tie build-up cavity, leading to $160$~mW at
531.2~nm. The second doubling uses an angle-tuned
90$^\mathrm{o}$-cut anti-reflection coated BBO crystal and delivers
up to $7$~mW at 265.6~nm. The DFB laser diode is injection-locked to
an ultra-stable laser source. As shown in fig.
\ref{fig:probe_light}, this laser source is composed of a Yb-doped
DFB fiber laser stabilized to an ultra-stable Fabry-P\'{e}rot cavity. A
fraction of the light is sent through an actively phase-stabilized
fiber link to stabilize an optical frequency comb generated by a
Ti:Sa femtosecond laser whose repetition rate is measured against
the LNE-SYRTE flywheel oscillator \cite{Chambon2005}, which is monitored by
several primary fountain frequency standards. Repeated measurements
of the ultra-stable laser source have shown highly predictable
behavior, relaxing the need for simultaneous operation of the
optical frequency comb with the rest of the experiment at the
current level of accuracy. Typically, measurements against the
primary frequency reference were performed several times a day which
is sufficient to estimate the optical frequency to better than
$100$~Hz at 1062.5~nm or 4 parts in $10^{13}$. Although this is
adequate for the present work, it is noteworthy that this probe
laser system already has the capability of highly stable and
accurate referencing limited only by primary frequency standards. In
fact, during the initial search for the clock transition, the
Yb-doped fiber laser was stabilized to a component of the optical
frequency comb, itself locked to the primary reference with a
controllable offset to allow for broader scanning of the probe
laser.

Spectroscopy of the clock transition is performed according to the
scheme proposed and demonstrated with strontium
\cite{Courtillot2003} and further used with ytterbium
\cite{Hoyt2005}. An up-going probe beam crosses the MOT at its
center and is either retro-reflected or not. Initially, the search
for the clock transition was performed with both the MOT and the
probe beam on continuously. Fluorescence of the MOT was monitored
as a function of the probe frequency.
Excitation of atoms in the untrapped, long-lived $^3P_0$ state
excited by the probe laser and their subsequent fall under gravity
induces losses in the MOT. Up to $90\%$ depletion of the MOT has
been observed. To suppress the light shift of $\sim 300$~kHz induced
by the 253.7~nm MOT beams and perform accurate measurements, the
following scheme is implemented. Using a mechanical shutter, the 2D-MOT and MOT beams are
switched on for an adjustable duration of $20$ to $50$~ms then off
for $5$~ms. During the 5 ms, the probe light is pulsed once with
adjustable delay and duration using acousto-optical modulator AOM1
in fig. \ref{fig:probe_light}. This cycle is repeated continuously.
During the $5$~ms, the atomic cloud is freely falling over
$122~\mu$m and expanding by $180~\mu$m at the Doppler temperature of
$31~\mu$K. Atoms are therefore recaptured when the MOT is turned
back on. The MOT reaches a steady state determined by the loading
rate during the MOT on time and the losses induced by the excitation
of the $^3P_0$ state. Average fluorescence of the MOT is measured
while stepping the probe frequency using AOM2 in fig.
\ref{fig:probe_light}.

\begin{figure}
\includegraphics[width= \linewidth]{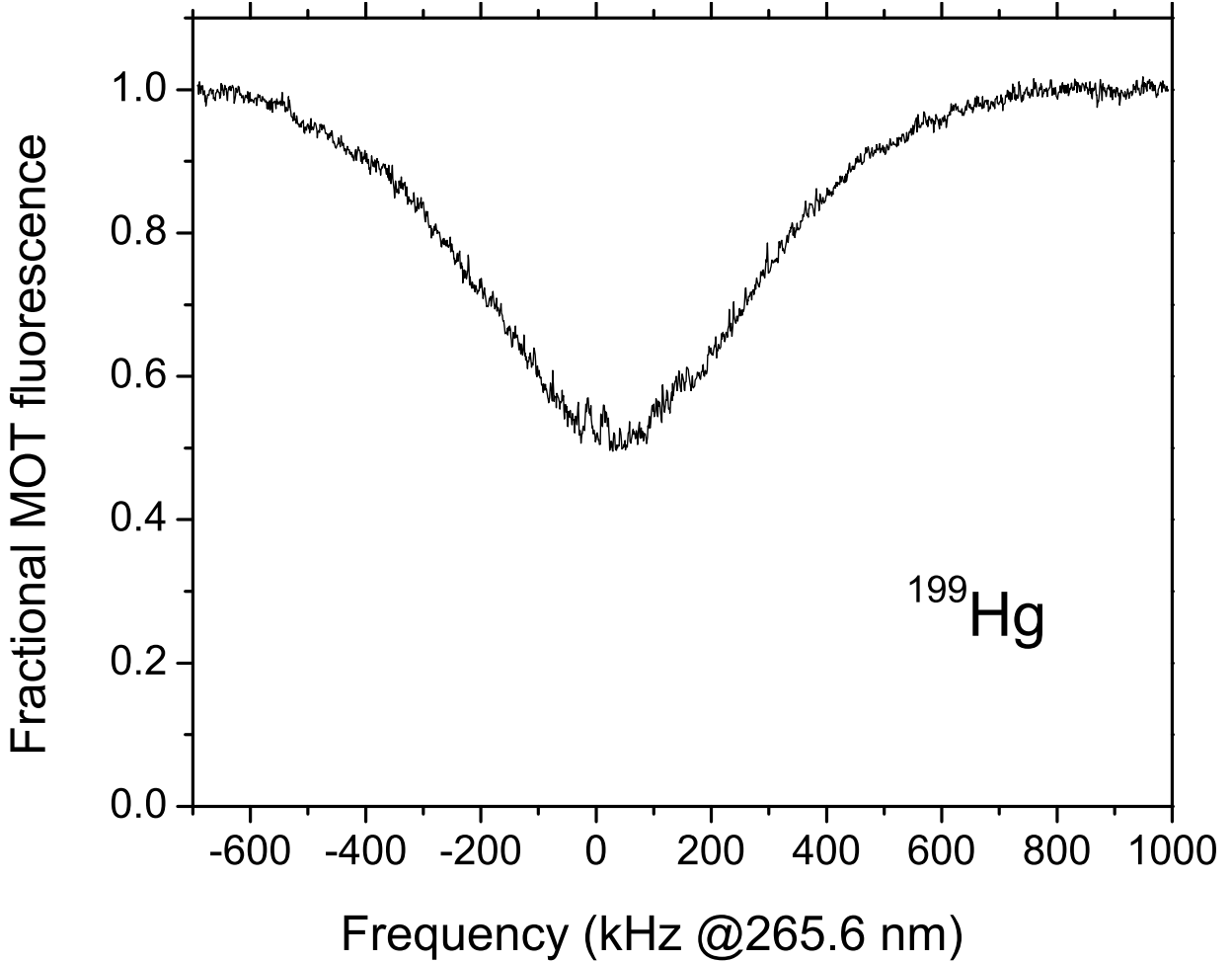}
\caption{\label{fig:spectra} Spectrum of $^1S_0-^3P_0$ transition in
$^{199}$Hg observed with a retro-reflected probe beam. Sharp
Doppler-free features can be seen at the top of the $580$~kHz wide
Doppler profile.}
\end{figure}


Measurements were first taken with a $1.4 \times 0.24~$mm up-going
probe beam. The maximum depletion of the MOT is $\sim 50\%$. The
observed peak is Doppler-broadened to a linewidth ranging from
$360$~kHz to $550$~kHz full-width at half maximum, depending on the
MOT parameters including the quality of beam alignment. The lowest
observed linewidth corresponds to a temperature of $36~\mu$K, which is close
to the Doppler temperature of $31~\mu$K related to the natural linewidth of the
cooling transition. Further measurements were performed with a
retro-reflected probe beam in order to cancel the Doppler shift
which is due to acceleration of the atoms under gravity and to a
possible non-vanishing average velocity of the cloud released from
the MOT. Comparisons of the frequency with single passed and
retro-reflected beams first led to estimating the uncertainty
related to the Doppler effect at $\sim 30-40$~kHz. Further
optimization of the retro-reflected geometry (size and overlap of
the returning beam) allowed for the observation of the Doppler-free
features expected for a retro-reflected probe. Figure
\ref{fig:spectra} shows a $^{199}$Hg spectrum measured with a probe
diameter of $280~\mu$m at $1/e^2$ and the maximum available power of
2~mW. Two sharp features can be distinguished at the top of the
Doppler profile which is $580$~kHz wide in this example.

\begin{figure}
\includegraphics[width= \linewidth]{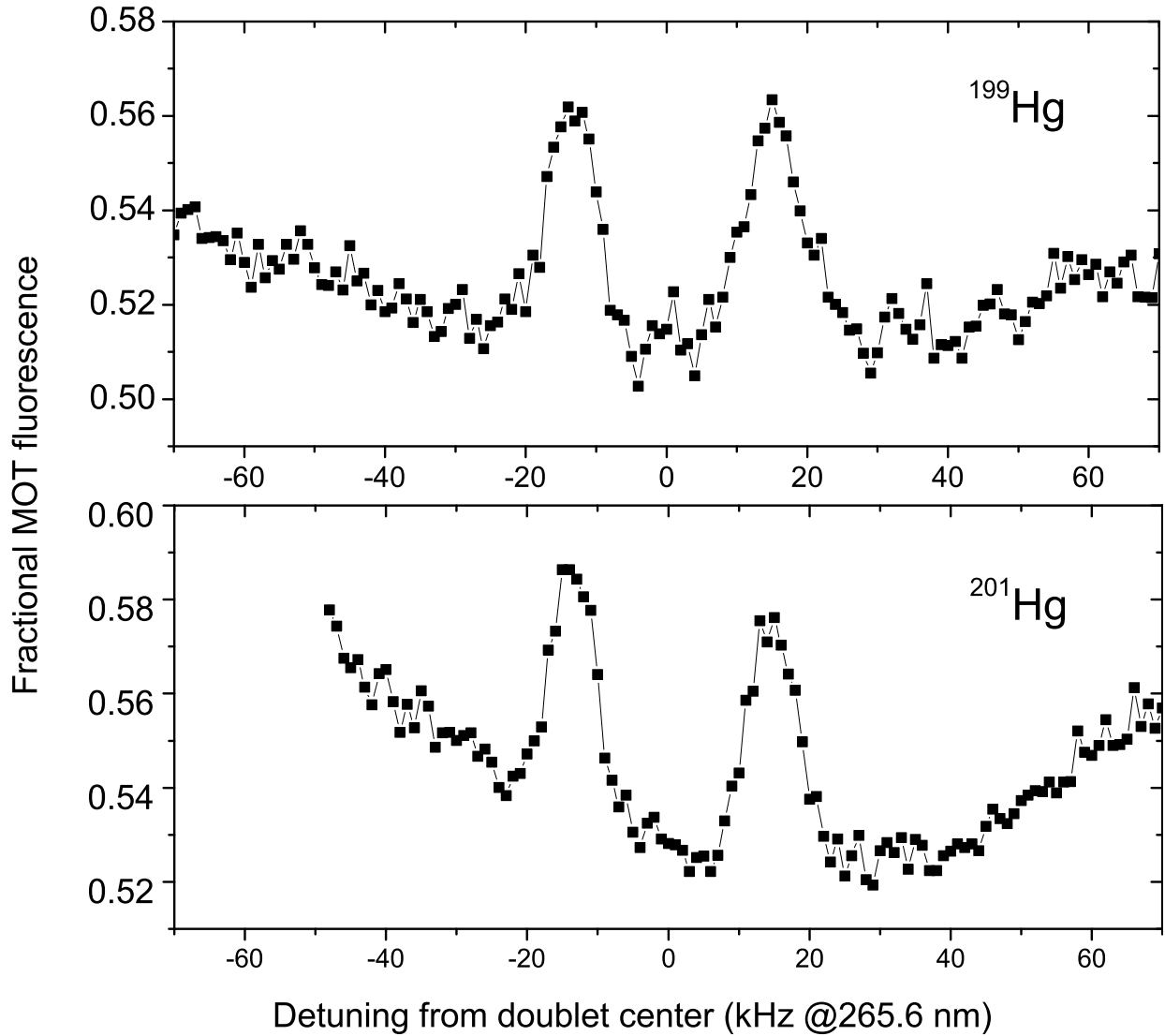}
\caption{\label{fig:recoildoublet} Recoil doublet observed in the
$^{199}$Hg (top) and $^{201}$Hg (bottom) spectra when a
retro-reflected probe beam is used. Data are taken at a rate of one point per second. Spectra have been averaged 4 times.}
\end{figure}

The two features, which are better seen on the narrower scans of
fig. \ref{fig:recoildoublet}, are the Doppler-free recoil doublet
\cite{Hall1976}. The recoil features are shifted by $\pm
\nu_{\mathrm{recoil}}$ with respect to the atomic transition
frequency $\nu$ , where $\nu_{\mathrm{recoil}}$ is the recoil
frequency of the transition equal to $\nu\times (h\nu/2 m
c^{2})$ or $14.2$~kHz, where $c$ is the speed of light, $h$ is
Planck's constant and $m$ is the atomic mass. The measured splitting
matches $2~\nu_{\mathrm{recoil}}$ within the overall statistical error bar of $\lesssim 1$~kHz.
We have also checked that the center of each component of the recoil
doublet is unchanged to less than $1$~kHz when the probe time is
changed by a factor of 2. Instead, under our experimental
conditions, the width of the recoil components is proportional to
the interaction time. Indeed, the maximum Rabi frequency that we
estimate based on the probe beam power and size, and from the
natural linewidth of the clock transition \cite{Bigeon1967} is $\sim
6$~kHz. It is smaller than the frequency chirp induced by the fall
under the acceleration due to gravity $g$ during the interaction time
$\tau$ which is $g \tau/\lambda \simeq 18$~kHz for a typical value
$\tau = 500~\mu$s. Note that the recoil doublet was also observed under similar conditions in  $^{40}$Ca
\cite{Fortier2006}. However, our $^{1}S_{0}$--$^{3}P_{0}$ mercury transition has a $3740$ times narrower natural linewidth.

To determine the atomic transition frequency, we take the center of
the recoil doublet. An uncertainty equal to the half-width at
half-maximum of the narrowest observed Doppler-free recoil feature
is conservatively assigned to this determination. This amounts to a
$4.5$~kHz uncertainty for both $^{199}$Hg and $^{201}$Hg.
When the MOT field is left on during the probe time, atoms at the
edge of the cloud can be exposed to a field of up to $0.2$~mT.
With the MOT field off, the residual field due to unshielded magnetic sources is
$\sim 0.3$~mT. A worst case estimation of the shift induced by the first order
Zeeman effect in such fields is $3.3$~kHz for $^{199}$Hg and $2.8$~kHz for $^{201}$Hg, given the nuclear magnetic moment
and the magnetic moment difference between the two clock states
\cite{Hachisu2008}. Measurements performed with and without
switching off the MOT field agree to within $700$~Hz. Ac Stark shift
of clock states due to the probe laser is less $100$~Hz.
Finally, it is noteworthy that the overall standard deviation of all measurements is less than $1$~kHz.
The measured frequencies are
$\nu(^{199}\mathrm{Hg})=1128575290808\pm 5.6$~kHz for $^{199}$Hg and
$\nu(^{201}\mathrm{Hg})=1128569561140\pm 5.3$~kHz for $^{201}$Hg. In
fractional terms, the uncertainty is $\sim 5$ parts in $10^{12}$ which
improves previous indirect determinations
\cite{saloman2006,Burns1952,Burns1952a} by more than $4$ orders of
magnitude. The isotope shift
between the two fermionic isotopes is $5729668\pm 7.7$~kHz. The
isotope shift between the best known bosonic isotope $^{198}$Hg
\cite{saloman2006} and $^{199}$Hg is
$\nu(^{198}\mathrm{Hg})-\nu(^{199}\mathrm{Hg})= 699\pm 12$~MHz with
an uncertainty dominated by the $^{198}$Hg uncertainty.

To summarize, we have reported the first laser-cooled spectroscopy of $^{1}S_{0}$--$^{3}P_{0}$ clock transition in fermionic isotopes of mercury.
Owing to the observation of the Doppler-free recoil doublet, we have measured
the transition frequency with an uncertainty which will make spectroscopy of the
clock transition in a lattice trap straightforward. This is an important step towards
producing a mercury lattice clock with unprecedented accuracy.


The authors would like to acknowledge support from SYRTE technical
services. SYRTE is Unit\'{e} Mixte de Recherche du CNRS (UMR CNRS 8630).
SYRTE is associated with Universit\'{e} Pierre et Marie Curie. This work
is partly funded by the cold atom network IFRAF. This work received
partial support from CNES.


\end{document}